\documentclass[11pt,twoside]{article}
\usepackage{FUSE2004}
\usepackage{natbib}

\usepackage{epsf}
\usepackage{psfig}
\usepackage{lscape}

\begin{document}
\title{The FUSE Survey of O~VI Absorption in the Galactic Disk }
\author{David V.~Bowen and Edward B.~Jenkins}
\affil{Princeton Observatory, Peyton Hall, Ivy Lane,
  Princeton, NJ~08544}
\author{Todd M.~Tripp}
\affil{Dept.\ of Astronomy, University of Massachusetts, Amherst,
  MA~01003}
\author{Kenneth R.~Sembach}
\affil{STScI, 3700 San Martin Dr.,
  Baltimore, MD~21218}
\author{Blair~D.~Savage}
\affil{University of Wisconsin, 475
N.~Charter Street, Madison, WI 53706}

\begin{abstract}

We outline the results from a FUSE Team program designed to
characterize O~VI absorption in the disk of the Milky Way.  We find
that O~VI absorption occurs throughout most of the Galactic plane, at
least out to several kpc from the Sun, and that it is distributed
smoothly enough for the column density to decline with height above
the disk and with distance in the plane. However, the O~VI absorbing
gas is clumpy, and moves at peculiar velocities relative to that
expected from Galactic rotation. We conclude that the observed
absorption is likely to be a direct indicator of the structures formed
when violent, dynamical processes heat the ISM, such as blowout from
multiple supernovae events.

\end{abstract}

\bigskip

\noindent The FUSE survey of O~VI absorption in the Galactic disk is a PI Team
program (program IDs P102 \& P122) designed specifically to
characterize O~VI absorption in the plane of the Galaxy. In selecting
suitable sightlines, we favored stars with distances $d \ga 1$ ~kpc,
with Galactic latitudes $|b|\la 10\deg$, and stellar types earlier
than B3.  Expending 720 ksec of PI time, and including additional data
from the FUSE Archive, we recorded O~VI absorption towards more than
150 stars.  To understand the distribution of O~VI in the Galaxy, we
combined our FUSE sample with data from the Copernicus satellite
(which largely sampled stars with $d \la 1$~kpc; see Jenkins
1978), as well as data from distant halo stars (Zsarg\'{o} et al.\ 2003),
and nearby white dwarfs (Oegerle et al.\ 2004). The analysis of these
data reveal the following:

\smallskip

\noindent{\bf 1.} Data from the ROSAT All-Sky Survey show that the
  positions of some stars coincide with regions of enhanced X-ray
  emission. These likely arise from hot bubbles blown out from the
  star or its association (Weaver et~al.\ 1977).  Many of our
  sightlines, however, lie at positions in the sky where there is
  apparently very little soft X-ray emission. For this subsample, we
  find that the average density along the sightline is slightly
  smaller than that measured towards all the stars; a simple estimate
  of the mid-plane density of O~VI,  $n_0$, for the subsample with no X-ray
  emission, derived from taking the median value of
  $N_i$(O~VI)/$d_i$ [as measured for each sightline $i$], yields
  $n_0=1.7\times 10^{-8}$~cm$^{-3}$.

\smallskip

\noindent{\bf 2.} We confirm the relationship between $N$(O~VI)$\sin
|b|$ and height above the Galactic plane $|z|$ found from FUSE
observations of extragalactic objects (Wakker et~al.\ 2003;
Savage~et~al.\ 2003). The correlation is conventionally characterized by
the relation $N$(O~VI)$\:\sin |b| = n_0 h [1-\exp ^{-|z|/h}]$, where
$h$ is the scale height of the O~VI above the plane. The value of
$n_0$ given above and $h\simeq 3.5$~kpc well fit the data, when
observations towards the extragalactic sightlines are included.

\smallskip

\noindent{\bf 3.} There is a correlation between distance in the
  Galactic plane and $N$(O~VI), over distances of $\approx 0.1-10$~kpc
  (Fig.~1). This shows that the processes which give rise to the
  existence of hot gas in the Galactic disk are ubiquitous over {\it
  all} the ISM between the outer spiral arms ($l=180\deg$) and the
  Galactic center ($l=0\deg$). The increase of $N$(O~VI) with distance
  also demonstrates that observed O~VI line profiles must be comprised
  of (perhaps many) blended components.

{\psfig
{figure=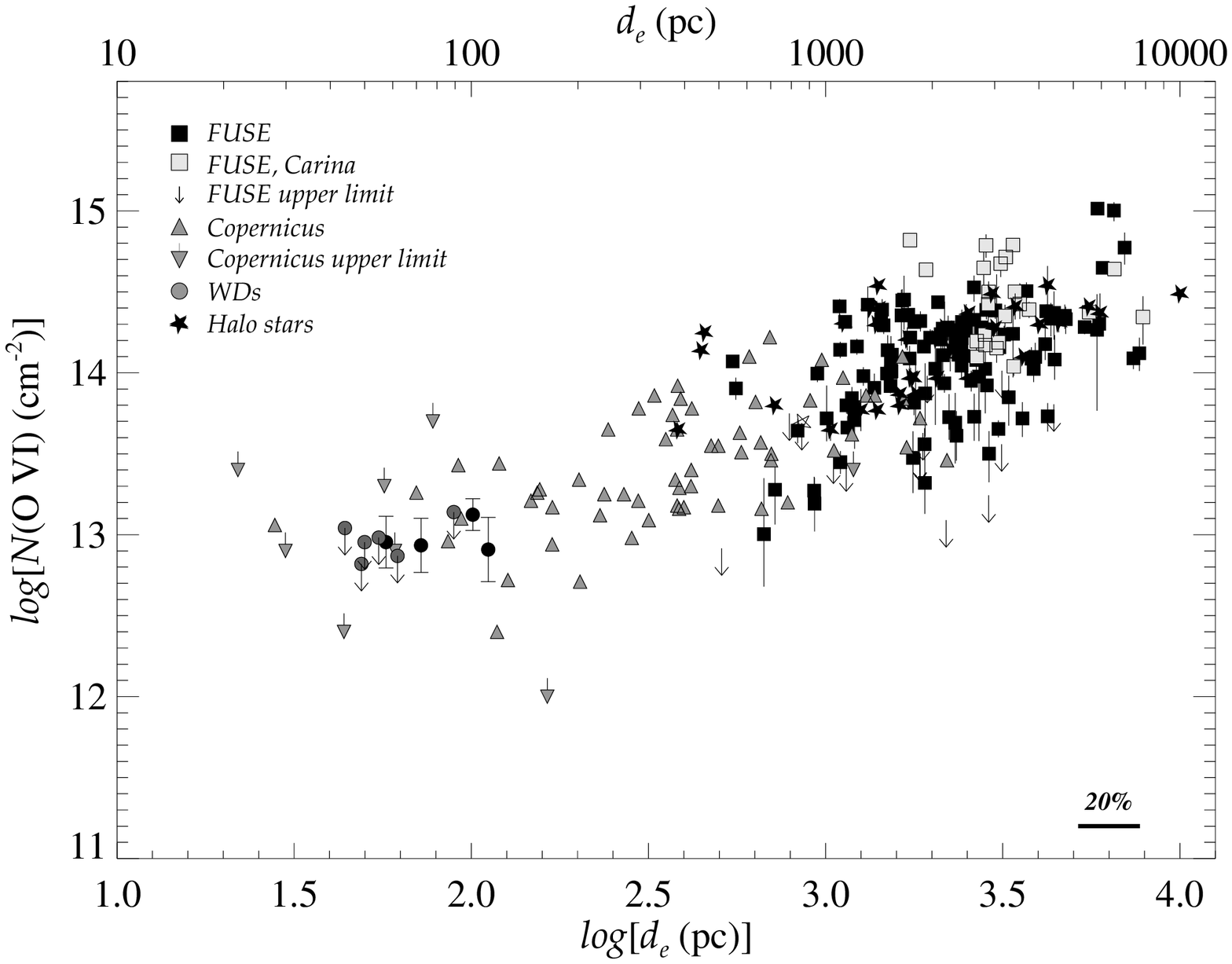,height=8cm,angle=90}}
\vspace*{-7cm}\hspace*{9.5cm}\parbox{3.1cm}{\scriptsize Fig.~1. Correlation of O~VI
  column density with the `effective' distance to a star $d_e$, which is the
  distance corrected for a decline in the O~VI density with $|z|$. 
  $d_e$ is given by  $h\csc |b|\,[1-\exp(-d\sin|b|/h)]$ (see
  Jenkins 1987). For these data, we use $h=3.5$~kpc. Although the
  error in the distance to a star depends on a number of factors, a
  representitive 20\% is shown at bottom right.}
\vspace*{1.6cm}

\noindent{\bf 4.} Significantly, although O~VI is distributed smoothly enough for
  $N$(O~VI) to correlate with $d$, the
  {\it dispersion} in the correlation does not decrease with distance, as
  would be expected by simply intercepting more
  clouds. For example, if a single cloud gave rise to a column density
  $N_0$, and the total observed column density increased as $n$ clouds
  were intercepted ($N=N_0\,n$), then we would predict the dispersion
  to go as $\pm\sqrt{n}\,N_0$. In fact, the fractional dispersion at large distances
  is similar to that seen at small distances. This 
  indicates that the O~VI is extremely clumpy compared to the
  distribution expected for either a smoothly distributed intercloud
  medium, or for randomly distributed uniform clouds.

\smallskip

\noindent{\bf 5.} We find a correlation between $N$(O~VI) and the
  width of O~VI absorption lines, as measured from their Doppler
  parameter, $b$. The fraction of oxygen in the form of O~VI is at its
  peak in collisionally ionized gas at temperatures of $\log T =5.5$
  (if in thermal equilibrium), when $1/4$ of the oxygen exists as
  O~VI. This corresponds to a Doppler width of $b=0.0321\sqrt{T} =
  17.6$~km~s$^{-1}$. Although the weakest lines in our sample have
  widths similar to this (at least within their errors), the majority
  of lines have widths considerably larger. The median value of all
  detected lines is $b=39.4$~km~s$^{-1}$, which would result in $\log
  T=6.2$, and $N$(O~VI)$\sim 1/400 N$(O), if the broadening was all
  thermal. The width of the lines cannot be accounted for by Galactic
  rotation; simulations of O~VI lines along the observed sightlines,
  assuming a smoothly distributed hot ISM, fail to reproduce the large
  Doppler widths measured in most cases.

\smallskip

\noindent{\bf 6.} Since  $N$(O~VI) correlates with $b$ and $d$, it is of no
  surprise to find that $b$ correlates with $d$, although not
  strongly. Since there is no reason to expect the temperature of the
  gas to increase with distance, this again demonstrates that
  the width of an O~VI line is determined by the velocity structure
  of overlapping components. The dispersion in $b$ as $d$
  increases is large, which may be the result of intercepting many
  different regions of hot gas each with very different line-of-sight
  velocities. 

\smallskip

\noindent{\bf 7.} We find that the velocities of the edges of O~VI
  absorption lines and those of strong C~II~$\lambda 1335$ and
  Si~III~$\lambda 1206$ lines (as measured in high resolution STIS
  data) are very similar. This suggests that the processes which move
  the hot gas to high velocity also affect small amounts of
  lower-ionization, cooler gas, in much the same way.

\smallskip

Interpretation of these results must therefore explain the fact that
O~VI production is common throughout the local Galactic disk, that it
arises in clumpy structures, and that it is dynamic. Of course, the
velocity of O~VI we measure need not be the true velocity of the gas;
for example, material which is flowing perpendicular to the disk will
have only a very small component of velocity along a line of sight. At
any rate, our data apparently rule out quiescent models of the hot
ISM: absorption cannot arise in a smooth, diffuse intercloud medium
which rotates along with the disk of the Milky Way.  Nor can O~VI arise in
`identical' hot clouds with similar column densities, randomly
disributed along a sightline---the hot ISM must be a much more
complicated structure than these simple models predict. A preliminary
analysis suggests that a population of clouds with sizes that follow a
power-law distribution may well produce the observed dispersion in
column densities. Our results are also consistent with recent
hydrodynamical simulations of the ISM, which predict that hot,
multiphase gas arises in turbulent, evolving structures as a result
of, e.g., multiple supernovae explosions, which produce overlapping
bubbles of shock-heated gas (see e.g.\ de Avillez \& Breitschwerdt in
this volume).  These detailed computer simulations are themselves
still evolving, and our observations will provide an important set of
results which these models can aim to reproduce.

\acknowledgements This research was funded by subcontract 2440$-$60014
from the Johns Hopkins University under NASA prime subcontract
NAS5-32985. 




\end{document}